\documentclass[%
 reprint,
 amsmath,amssymb,
 aps,
floatfix,
]{revtex4-2}

\usepackage{graphicx}
\usepackage{caption}
\usepackage{subcaption}
\usepackage{dcolumn}
\usepackage{bm}
\usepackage{cleveref}

\font\tenscr=rsfs10 scaled1100
\font\sevenscr=rsfs7 
\font\fivescr=rsfs5 
\skewchar\tenscr='177
\skewchar\sevenscr='177
\skewchar\fivescr='177
\newfam\scrfam
\textfont\scrfam=\tenscr
\scriptfont\scrfam=\sevenscr
\scriptscriptfont\scrfam=\fivescr

\def\scri{{\fam\scrfam I}}

\begin{document}

\title{Numerical evolution of the center of mass and angular momentum for binaries black holes}

\author{Emmanuel A. Tassone }
\email{emmanuel.tassone@unc.edu.ar}
\author{Paula A. Mandrilli, Carlos N. Kozameh, Gonzalo D. Quiroga, and José I. Nieva}

\affiliation{
 FaMAF, Universidad Nacional de C\'ordoba\\
\normalsize \em \small 5000, C\'ordoba, Argentina 
}

\date{\today}

\begin{abstract}
The asymptotic approach derived by Kozameh-Quiroga (K-Q) provides a modern framework
to obtain the evolution of global variables of isolated sources of gravitational radiation. We test
the K-Q formalism evolving the equations of motion for the center of mass, the intrinsic angular
momentum, and several other global variables, for black hole binary coalescence. First we
evolve the equations of motion using 777 simulations from the RIT catalogue numerical data of $\psi_4$ \cite{healy2020}. We then analyze the trajectory of the center of mass and compute the final state of other physical variables after the coalescence has taken place. Finally, we show the results obtained from our equations of motion are consistent with those in the Rochester metadata.

\end{abstract}

\maketitle


\section{Introduction}\label{section1}

The equations of motion for global variables describing isolated sources of gravitational radiation is an old and important area of research in general relativity (GR). Out from the many formulations presented in the literature, those using the mathematical formulation of asymptotic flatness offer the possibility of matching its predictions with observations.

The notions of Bondi mass, linear momentum, dipole mass moment and angular momentum resemble their minkowskian counterparts in classical field theory, except that in this case they are not conserved quantities but obey evolution equations that are derived from the field equations of GR. Linking those quantities with equations of motion for the center of mass and intrinsic angular momentum is the goal of the different approaches given in the literature \cite{geroch1981,moreschi2004,szabados2009,flanagan2017,helfer2007}.

Many of the key ideas to define and evolve those variables were presented and developed by E.T. Newman and collaborators\cite{adamo2012,kozameh2016}. More recently, these results were generalized and a relativistic approach is now available in the literature \cite{kozameh2020}.

Since binary coalescence offers an excellent scenario to test predictions both in terms of numerical solutions, semianalytical approaches and direct observation of gravitational radiation, the purpose of this work is to numerically evolve the equations of motion for the center of mass and intrinsic angular momentum for some 800 cases that were taken from the Rochester repository \cite{healy2020}.

The main idea in this work is to present the equations, give technical details of the numerical setup and show some results both for particular cases and for the whole numerical evolutions.

To classify the binary system type, we follow the conventions of the Rochester repository, dividing the numerical outcomes in different classes according to the relative masses into equal or nonequal masses (EM or NEM). Likewise, the spins of the black holes are taken into account and we can distinguish them between nonspinning (NS), Aligned with the orbital angular momentum (A) or precessing spins (P).

In Section 2 we give a brief outline of the definition of these global variables and their equations of motion and offer the reader the relevant references with thorough approaches.

In Section 3 we discuss the technical issues of the numerical setup. The relationship between spin weighted and tensorial spherical harmonics is used to link the gravitational radiation of the repository with the equations of motion for the global variables. 

In Section 4 we present some results that show the global behavior of these classes of binaries. The center of mass trajectories for selected members of the 6 classes are given, the tilt and change of the angular momenta are obtained for all cases, kickback velocities are also computed. The results obtained from our equations of motion are consistent with those in the Rochester metadata.

\section{Equations of motion for the center of mass and intrinsic angular momentum} \label{section2}
There are many results that are needed for this work. In this section, we introduce several of the key ideas and the basic tools that are useful for our later discussion.

\subsection{Asymptotically flatness and $\scri^+$}

We assume our spacetime is empty except for a bounded region where the dynamical interaction of the sources of gravitational radiation takes place. For such spacetimes there is a well known mathematical structure where a null boundary $\scri^+$ is added together with an appropriate definition of global variables that take into account the time evolution of the sources. In a neighborhood of $\scri^+$ the gravitational field is given by the Weyl tensor. In the neighborhood of null infinity, it is also possible to introduce a particular system called Bondi system. A Bondi system is an inertial frame in general relativity, whose coordinates are labeled by $(u_B,r_B,\zeta_B,\bar{\zeta}_B)$. The time $u_B$ represent null surfaces, $r_B$ is the affine parameter along the null geodesics of the constant $u_B$ surfaces and $\zeta_B,\bar{\zeta}_B$ are the complex stereographic coordinates.

Using a null tetrad adapted to a Bondi system one defines five complex scalars, whose asymptotic behavior is \cite{held1980}
\begin{eqnarray*}
	\psi_{0} &=&{C_{abc}}^dm^{a}l^{b}l^{c}m_{d}\simeq \frac{\psi _{0}^{0}}{r_B^{5}}, \quad \psi _{3} ={C_{abc}}^dl^{a}n^{b}n^{c}\bar{m}_{d}\simeq \frac{\psi _{3}^{0}}{r_B^{2}},\\
	\psi_{1} &=&{C_{abc}}^dn^{a}l^{b}l^{c}m_{d}\simeq \frac{\psi _{1}^{0}}{r_B^{4}}, \quad \psi _{4} ={C_{abc}}^d\bar{m}^{a}n^{b}n^{c}\bar{m}_{d}\simeq \frac{\psi_{4}^{0}}{r_B},\\
	&& \psi _{2} =\frac{1}{2}({C_{abc}}^dl^{a}n^{b}m^{c}\bar{m}_{d}-C_{abcd}l^{a}n^{b}l^{c}n_{d})\simeq\frac{\psi _{2}^{0}}{r_B^{3}}.
\end{eqnarray*}

With help of the peeling theorem, which describes the asymptotic behavior of the Weyl tensor, the radial part of the Einstein equations can be integrated leaving only the Bianchi identities at $\scri^+$ as the unsolved equations. In a Bondi frame the resulting equations look remarkably simple. Some of those equations relate the Weyl scalars with the Bondi shear, i.e., \cite{adamo2012,held1980}
\begin{eqnarray}
	\psi_{2}^{0}+\eth^{2}\bar{\sigma }^{0}+\sigma ^{0}\dot{\bar{\sigma }}^{0}&=&\bar{\psi }_{2}^{0}+\bar{\eth }^{2}\sigma ^{0}+\bar{\sigma }^{0}\dot{\sigma }^{0},\label {psi2}\\
	\psi_{3}^{0}&=&\eth \dot{\bar{\sigma}}^0,\\
	\psi_4^{0}&=&-\ddot{\bar{\sigma}}^0 .\label{ddot_psi4}
\end{eqnarray}
Here the operator $\eth$ is known as the ``eth operator'' and is basically the complex covariant derivative over $S^2$ in stereographic coordinates. The derivative is taken at $u_B=const$ surfaces \cite{goldberg1967}.

Finally, the Bianchi identities (in Bondi coordinates) at $\scri^+$ are given
by \cite{adamo2012,held1980}
\begin{align}
	\dot{\psi}_{0}^{0}&=-\eth \psi_1^0+3\sigma ^{0}\psi_2^0,\\
	\dot{\psi}_{1}^{0}&=-\eth \psi_2^0+2\sigma ^{0}\psi_3^0,\\
	\dot{\psi}_2^0&=-\eth \psi_3^0 +\sigma^0\psi_4^0.\label{Psi_2dot}
\end{align}
Note that eq. (\ref{psi2}) defines a real variable $\Psi$ called the mass aspect \cite{kozameh2018}.
\begin{equation} \label{asp.masa}
	\Psi =\psi _{2}^{0}+\eth ^{2}\bar{\sigma }^{0}+\sigma ^{0}\dot{\bar{\sigma }}^{0}.
\end{equation}
In term of $\Psi$ is possible to write the Bondi Mass $M$ and Bondi lineal momentum $P^i$ by
\begin{eqnarray}
	M &=&-\frac{c^{2}}{8\pi \sqrt{2}G}\int \Psi dS,\\
	P^{i} &=&-\frac{c^{3}}{8\pi \sqrt{2}G}\int {\Psi }\tilde{l}^{i} dS,
\end{eqnarray}
with
\begin{equation}
	\tilde{l}^{i}=\frac{1}{1+\zeta \bar{\zeta}}(\zeta +\bar{\zeta},-i(\zeta -%
	\bar{\zeta}),1-\zeta \bar{\zeta}),
\end{equation}
$dS=\frac{4d\zeta \wedge d\bar\zeta}{P_0^2}$ the area element on the unit sphere and where $i,j,k,l,m=1,2,3$ are three dimensional Euclidean indices. 

It is also quite convenient to give the evolution equation for $\Psi$. Directly from (\ref{Psi_2dot}) one obtains
\begin{equation}
	\dot{\Psi}=\dot{\sigma} ^{0}\dot{\bar{\sigma }}^{0}.\label{psiprima}
\end{equation}
The above equation or (\ref{Psi_2dot}) contain identical information.

\subsection{Center of mass and intrinsic angular momentum}
The center of mass and intrinsic angular momentum are useful concepts for isolated systems in General Relativity. One can obtain global features from binary coalescence such as the kickback velocity, reaction radiation force, angular momentum loss, among other physical quantities. Following a formulation presented several years ago \cite{kozameh2016}, to  obtain such variables, we first introduce special congruences of generalized Newman-Unti (NU) cuts at null infinity $$u_B=Z(x^a(u),\zeta,\bar{\zeta}),$$ 
with $x^a(u)$ an arbitrary worldline in a fiducial Minkowski space. The above equation has a geometrical kinematical meaning. It represents the intersection of the future light cone of each point $x^a(u)$ with Scri and it is called a null cone cut congruence. In general, these cuts have caustics but it is possible to construct the so called regularized null cone cut congruences with a well defined procedure that yields regular cuts at null infinity. In flat spacetime, the regularized cuts are given by
\begin{eqnarray}
Z_{0}=t(u)-\frac{1}{2}x^{i}(u)Y_{1i}^{0} \label{Z0}
\end{eqnarray}
where $Y_0^0,Y^0_{1i},Y^{0}_{2ij}$ are the tensorial spin-s harmonic \cite{newman2005}, and $x^a(u)$ an arbitrary worldline in Minkowski spacetime. The linearized deviation from flat space is given by \cite{kozameh2020},
\begin{eqnarray} \label{Z1}
Z_{1}=Z_{0}+\left( \frac{\Delta\sigma _{R}^{ij}}{12}+\frac{1}{72}\dot{{\sigma}}_{I}^{ig}x(u)^{f}\epsilon ^{gfi}\right) Y_{2ij}^{0}, \label{Z1}\nonumber\\
\end{eqnarray}
where the extra term is the quadrupolar contribution of the Bondi shear with $\Delta\sigma _{R}^{ij}$  the deviation from the initial real part of the shear. $\dot{{\sigma}}_{I}^{ig}$ is the derivative with respect to Bondi time of the shear imaginary part and $\epsilon^{gfi}$ is the Levi-Civita symbol. The extra term in \ref{Z1} explicitly depends on the gravitational radiation and it vanishes for a stationary situation. If $x^a(u)$ describes any worldline, then $Z_{1}$ describes a NU foliation up to the order needed for this calculation. From the geometrical meaning outlined before, the above equation represents NU congruences parametrized by worldlines on a Minkowski spacetime. This fiducial flat space will be used to write down the equations of motion for the center of mass and intrinsic angular momentum.

We then use the Winicour linkages \cite{held1980:3} to define the notion of dipole mass moment and angular momentum on these null cone cut congruences from the real and imaginary parts of the linkage integral as follows\cite{kozameh2016,kozameh2018},
\begin{eqnarray}\label{DJ}
	D^{\ast i} + i\ c^{-1}J^{\ast i}=-\frac{c^{2}}{12\sqrt{2}G}\left[ \frac{2\psi _{1}^{0}-2\sigma
		^{0}\eth \bar{\sigma}^{0}-\eth(\sigma ^{0}\bar{\sigma}^{0})}{Z^{\prime 3}}\right]^{\ast i}. \nonumber\\
\end{eqnarray}
The reader should here distinguish between the $i$ labeling the index of the equation and the $i$ multiplying the factor $c^{-1}$ , which is the imaginary unit. This latter $i$ will no longer appear after we separate the equation into its real and imaginary part.

Finally, we assume that among all possible null cone cut congruences, there exists a special world line $R^a(u)$ such that at each $u=const.$ cut the mass dipole moment $D^{\ast i}$ vanishes. This special world line will be called the center of mass world line of the system. The angular momentum $J^{i\ast}$ evaluated at the center of mass will be the intrinsic angular momentum $S^i$. 

From the point of view of Bondi observers for each $u=const.$ cut, there exist a Lorentz boost and a translation that relate the above defined variables to the equivalent definitions given on Bondi cuts, i.e., if we define $D^{i}$ and $J^{i}$ in a Bondi system as, 
\begin{equation}\label{DJB}
	D^{i} + ic^{-1}J^{i}=-\frac{c^{2}}{12\sqrt{2}G}\left[ 2\psi _{1}^{0}-2\sigma
	^{0}\eth \bar{\sigma}^{0}-\eth(\sigma ^{0}\bar{\sigma}^{0})\right]^{i}, 
\end{equation}
then the transformation law between the quantities $(\psi _{1}^{0\ast },\sigma^{0\ast}, \eth)$ and 
$(\psi _{1}^{0}, \sigma^{0}, \eth_{B})$ 
together with the condition that in the center of mass foliation the dipole mass moment vanishes, i.e. $D^{i\ast}|_{u=const}=0$, yields the relativistic definition of center of mass world line. The details can be found in reference \cite{kozameh2016,kozameh2018}. The final results read,

\begin{eqnarray}
	D^{i}&=&MR^{i}+\frac{1}{c^2}\epsilon^{ijk}V^{j}S^{k}-\frac{8}{5\sqrt{2}c} P^{j}\Delta\sigma _{R}^{ij}\nonumber\\
	&&-\frac{c^2}{G}\epsilon^{ijk}(\frac{4}{5}\sigma_I^{jl}\sigma_R^{kl}-\frac{36}{7}\sigma_I^{klm}\sigma_R^{jlm}),\\
	J^{i}&=&S^{i} + \epsilon^{ijk}R^{j}P^{k}\nonumber\\
	&&-\frac{151c^2}{168\sqrt{2}G}(\sigma_R^{ijk}\sigma_I^{jk}-\sigma_I^{ijk}\sigma_R^{jk}),
\end{eqnarray}
where the above equations have been written keeping up to linear terms in the velocity and up to second order in the radiation fields. We have only kept quadrupole and octupole terms in the Bondi shear since these terms capture the main part of the gravitational radiation given in the repository. 

The main contribution to the gravitational radiation comes from the $l=2$ and $l=3$ spherical harmonic decomposition.

The above equations relate the center of mass $R^{i}$ and intrinsic angular momentum $S^{j}$ with the asymptotic variables $D^{i}$, $M$, $P^{i}$ and $J^{i}$, given on a Bondi coordinate system. Note also that $V^i=\frac{1}{\sqrt{2}}\dot{R}^{i}$ since the relativistic time is $t=u\sqrt{2}$ .

We list in \cref{Dpunto,Jpunto,Mpunto,Ppunto} the evolution equations of the Bondi physical variables that are needed for this work. The evolution equations of $D^i$ and $J^i$ follow from the Bianchi identity for $\psi_1^0$ when the $\ell=1$ component of the real and imaginary parts of $\dot {\psi}_1^{0}$ is computed \cite{kozameh2020}. Furthermore, the dynamical evolution of the Bondi mass $M$ and momentum $P^{i}$ can be computed from the Bianchi identity for $\dot{\psi}_2^0$. These equations are given by, 

\begin{align}
	\dot{D}^{i}&=\sqrt{2}P^{i}+\frac{3}{7}\frac{c^2}{\sqrt{2}G}\big[(\dot\sigma_R^{ijk}\sigma_R^{jk}-\sigma_R^{ijk}\dot\sigma_R^{jk}) \big] \nonumber\\
	&+\frac{3}{7}\frac{c^2}{\sqrt{2}G}\big[(\dot\sigma_I^{ijk}\sigma_I^{jk}-\sigma_I^{ijk}\dot\sigma_I^{jk}) \big] ,\label{Dpunto}\\
	\dot{J}^{i}&=\frac{c^{3}}{5G}(\sigma _{R}^{kl}\dot{\sigma}_{R}^{jl}+\sigma _{I}^{kl}\dot{\sigma}_{I}^{jl})\epsilon^{ijk} \nonumber\\
	&+\frac{9c^3}{7G}(\sigma_R^{klm}\dot\sigma_R^{jlm}+\sigma_I^{klm}\dot\sigma_I^{jlm})\epsilon^{ijk},\label{Jpunto}\\
	\dot{M}&=-\frac{c}{10\sqrt{2}G}(\dot{\sigma}_{R}^{ij}\dot{\sigma}_{R}^{ij}+\dot{\sigma}_{I}^{ij}\dot{\sigma}_{I}^{ij})\nonumber\\
	&-\frac{3c}{7\sqrt{2}G}(\dot\sigma_R^{ijk}\dot\sigma_R^{ijk}+\dot\sigma_I^{ijk}\dot\sigma_I^{ijk}  ),\label{Mpunto}\\
	\dot{P}^{i}&=\frac{2c^{2}}{15\sqrt{2}G}\dot{\sigma}_{R}^{jl}\dot{\sigma}_{I}^{kl}\epsilon^{ijk}-\frac{\sqrt{2}c^2}{7\sqrt{2}G}(\dot\sigma_R^{jk}\dot\sigma_R^{ijk}+\dot\sigma_I^{jk}\dot\sigma_I^{ijk})\nonumber\\
	&+\frac{3c^2}{7\sqrt{2}G}\dot\sigma_R^{jlm}\dot\sigma_I^{klm}\epsilon_{ijk}.\label{Ppunto}
\end{align}

The main idea then is to give initial data to solve for   \cref{Dpunto,Jpunto,Mpunto,Ppunto}, assuming that before the gravitational radiation is emitted the center of mass world line is the origin of a Bondi coordinate system. Once the Bianchi identities have been solved we introduce a perturbation procedure that preserves the order of the equations and algebraically solve for $R^{i}$ and $S^{j}$. 

It is worth noting that the equations of motion are written in terms of the Bondi shear whereas the gravitational radiation is given by $\psi^0_4$. To find the relationship between those scalars we first write  \cite{nakano2015} 
\begin{equation}
\psi^0_4=\sum\limits_{l,m}\Psi^{lm}{}_{(-2)}Y_{lm}.
\end{equation}
Here $\Psi^{lm}$ represents the multipole of the gravitational wave at infinity. In practice, $\Psi^{lm}$ are functions stored in data files generated by the numerical simulation of the astrophysical systems such as the binary coalescence of Black Holes \cite{healy2014}.

We then use equation (\ref{ddot_psi4}) to write
\begin{align} \label{link}
\ddot{\bar{\sigma}}^0=-\sum\limits_{l,m}\Psi^{lm}{}_{(-2)}Y_{lm}.
\end{align}

Integrating twice the above equation one obtains the real and imaginary parts of the Bondi shear. Usually one assumes the real part of the shear vanishes initially but it is just a matter of convenience since the equations are supertraslation invariant.
Also, in many works in GR, it is usual to expand the shear, and other scalars, in terms of the tensorial harmonics as follows (cf. \cite{newman2005}), 
\begin{align}
\sigma^0&= \sigma^{ij}Y_{2ij}^2+\sigma^{ijk}Y_{3ijk}^2+...\\
\ddot{\bar{\sigma}}^0&= \ddot{\bar{\sigma}}^{ij}Y_{2ij}^{-2}+\ddot{\bar{\sigma}}^{ijk}Y_{3ijk}^{-2}+...
\end{align}
where $\sigma^{ij}$ and $\sigma^{ijk}$ are related to the quadrupole and octupole contribution of the gravitational wave respectively. On the other hand, the spin weighted spherical harmonic decomposition is often used in the literature of gravitational radiation. Thus, to be able to find the evolution of $R^{i}$ and $S^{j}$ one must first obtain a correspondence between $Y_{2ij}^{-2} \rightarrow {}_{(-2)}Y_{2m} $  and $Y_{2ijk}^{-2} \rightarrow {}_{(-2)}Y_{3m} $ (see \cite{mandrilli2020}) and then solve for $\ddot{\bar{\sigma}}^{ij}$. This is  further discussed in the next section. The relevant components of the shear needed for this work are given by,
\begin{align}
\ddot{\bar\sigma}^{xy}&=-\frac{i}{4}\sqrt{\frac{5}{\pi}}(\Psi^{22}-\Psi^{2-2}),\\
\ddot{\bar\sigma}^{xx}&=-\frac{1}{4}\sqrt{\frac{5}{\pi}}(\Psi^{2-2}+\Psi^{22})+\frac{1}{6}\sqrt{\frac{15}{2\pi}}\Psi^{20},\\
\ddot{\bar\sigma}^{yy}&=\frac{1}{4}\sqrt{\frac{5}{\pi}}(\Psi^{2-2}+\Psi^{22})+\frac{1}{6}\sqrt{\frac{15}{2\pi}}\Psi^{20}, \\
\ddot{\bar\sigma}^{yz}&=\frac{i}{4}\sqrt{\frac{5}{\pi}}(\Psi^{2-1}+\Psi^{21}),\\
\ddot{\bar\sigma}^{xz}&=\frac{1}{4}\sqrt{\frac{5}{\pi}}(\Psi^{21}-\Psi^{2-1}),
\end{align}
and 
\begin{align}
\ddot{\bar{\sigma}}^{xxx}&=\frac{1}{8}\sqrt{\frac{7}{5\pi}}(\Psi^{3-1}-\Psi^{31})+\frac{1}{8}\sqrt{\frac{7}{3\pi}}(\Psi^{33}-\Psi^{3-3})\\
\ddot{\bar{\sigma}}^{xyy}&=\frac{1}{24}\sqrt{\frac{7}{5\pi}}(\Psi^{3-1}-\Psi^{31})-\frac{1}{8}\sqrt{\frac{7}{3\pi}}(\Psi^{33}-\Psi^{3-3})\\
\ddot{\bar{\sigma}}^{yyy}&=-\frac{i}{8}\sqrt{\frac{7}{5\pi}}(\Psi^{3-1}+\Psi^{31})-\frac{i}{8}\sqrt{\frac{7}{5\pi}}(\Psi^{33}+\Psi^{3-3})\\
\ddot{\bar{\sigma}}^{xxy}&=-\frac{i}{24}\sqrt{\frac{7}{5\pi}}(\Psi^{3-1}+\Psi^{31})+\frac{i}{8}\sqrt{\frac{7}{5\pi}}(\Psi^{33}-\Psi^{3-3})\\
\ddot{\bar{\sigma}}^{xxz}&=-\frac{1}{12}\sqrt{\frac{7}{2\pi}}(\Psi^{32}+\Psi^{3-2})+\frac{1}{4}\sqrt{\frac{7}{15\pi}}\Psi^{30}\\
\ddot{\bar{\sigma}}^{yyz}&=\frac{1}{12}\sqrt{\frac{7}{2\pi}}(\Psi^{32}+\Psi^{3-2})+\frac{1}{4}\sqrt{\frac{7}{15\pi}}\Psi^{30}\\
\ddot{\bar{\sigma}}^{xyz}&=\frac{i}{12}\sqrt{\frac{7}{2\pi}}(\Psi^{3-2}-\Psi^{32}).
\end{align}

\section{The numerical evolution}\label{section3}
In this section we explain how the time evolution is made. The idea is basically to obtain the sigma tensors in which our formalism is described. We thus use the set of equations derived from the Bianchi identities at null infinity presented in \ref{Dpunto}-\ref{Ppunto} . They are valid in any Bondi frame and can be numerically integrated to obtain their time evolution. 

\subsection{A Dictionary between tensorial and spin weighted spherical harmonics}
The solutions of equations \ref{Dpunto}-\ref{Ppunto} are then used to algebraically solve for the center of mass $R^i$ and intrinsic angular momentum $S^i$.
In order to obtain the mass dipole moment,  the angular momentum and the linear momentum, we need only the $2$ and $3$ modes  of $\Psi^{lm}$ (taken from \cite{RITcatalog}) to get the values of the shear components shown above. Higher order modes correspond with higher order tensor decomposition of sigmas and thus, $l=4$ modes are irrelevant for the precision order to which our formulation applies.

As a first step, we establish a relationship between the spin weighted spherical harmonics used in the catalog and the tensorial spherical harmonics used in the definitions of R and S.

To expand the Weyl scalar $\psi_4$ of the Newman-Penrose formalism \cite{adamo2012,kozameh2016,kozameh2018} and to map the tensorial Bondi shear into the multipole associated with the gravitational wave we use the one-to-one correspondence between tensorial spin-s harmonics denoted by $Y^{(s)}_{(l)i...k}$ \cite{newman2005} and the spin-weighted spherical harmonics ${}_{(s)}Y_{lm}$. This correspondence was shown in \cite{mandrilli2020}.

\subsection{Sigma calculus}\label{SigmaCalculus}
Obtaining the Bondi shear implies integrating twice the scalar $\psi_4$. There is a
well-known nonlinear drift which appears as a secular effect coming from the double integration \cite{berti2007,reisswig2011}. On top of this there is also a linear drift coming from the initial values of $\sigma_B$ and $\dot{\sigma}_B$ that must be given in the integration. These drifts must be corrected in order to obtain a more accurate shear and thus more accurate final parameters of the evolution.

Reisswig \cite{reisswig2011} has implemented a code to deal with those problems performing the Fourier transformation with a windowing function and a low-frequency cutoff (which should be lower than all physically possible frequencies) \cite{reisswigcode}. The code provides the strain of the gravitational radiation, commonly denoted in the literature as $h$.

\begin{figure}[htp]
     \begin{subfigure}{0.5\textwidth}
         \centering
         \includegraphics[width=\textwidth]{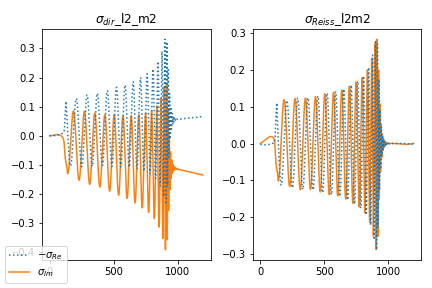}
         \caption{Comparison of modes l=2,m=2.}
         \label{fig0:f1}
     \end{subfigure}\hfill
     \begin{subfigure}{0.5\textwidth}
         \centering
         \includegraphics[width=\textwidth]{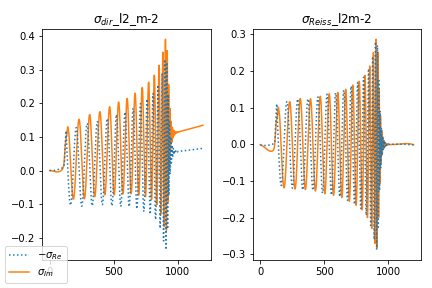}
         \caption{Comparison of modes l=2,m=-2.}
         \label{fig0:f2}
     \end{subfigure}
     \caption{Integration of the modes which made the most important contribution to simulation 0014. On the left $\sigma_{dir}$ are the plots of integrating directly . On the right, $\sigma_{Reiss}$ are the sigmas obtained when Reiss code is implemented. }
\end{figure}

For the purpose of this work, which is to check qualitatively the consistency of the K-Q formalism, it is understandable to disregard the drifts for a first overview of the formalism.
 In  \cref{fig0:f1,fig0:f2} we show the difference from integrating with Reisswig code (blue dotted line) and making an integration directly (orange solid line). Note the plot is in terms of the quantity $\sigma$ which is not directly obtained by the code but is easy to obtain with the relation $h=-\sigma$, where $h$ is the strain and $\sigma$ the shear we use along all this writing.  The drifts produce a change of order $\thicksim 0.2$ at most (largest component $\sigma$). Moreover, \cref{Ppunto,Dpunto,Jpunto,Mpunto} are quadratic on $\sigma_{ij}$ hence making the drift effect smaller ($\thicksim 0.04$). This will be the case for the majority of simulations also. The incorporation of this effect is left for a future and more accurate study.
\subsection{Getting rid of traveling waves}
Another problem we address is how to get rid of traveling waves that are present on any initial data set.
We take as an example the evolution of the 0443 simulation whose gravitational radiation is plotted below. We plot the square root of the integral of $\Psi_4^0 \bar{\Psi_4^0}$ on the sphere to take into account the different contributions of the $l=2$ and $l=3$ terms.
\begin{figure}[h]
	\centering
	\includegraphics[scale=0.5]{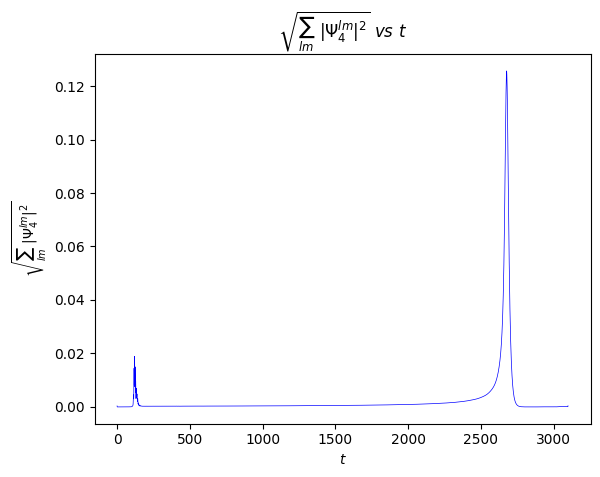}
	\caption{Evolution of the gravitational radiation for the simulation case 0443}
	\label{fig1}
\end{figure}
 One can see a traveling wave that is not relevant for the coalescence problem at the beginning of the time evolution. This will produce an unwanted jump of the Bondi linear momentum.
 
\begin{figure}[h]
	\centering
	\includegraphics[scale=0.6]{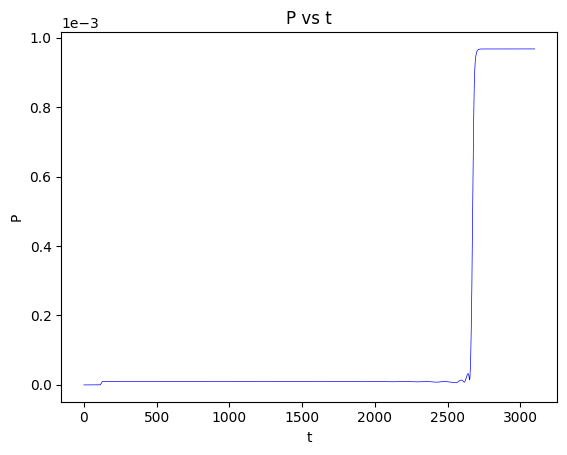}
	\caption{Time evolution of the magnitude of the Bondi momentum. A small change between 0 and 500 due to the traveling wave can be seen.}
	\label{fig2}
\end{figure}

The plot shows the evolution of the Bondi momentum using the Rochester repository on a Bondi frame whose origin is at the center of mass which is initially at rest. After the traveling wave passes by, the momentum is no longer zero and remains constant until the gravitational wave of the binary system is emitted. After that the Bondi momentum is constant and non vanishing.
Since a traveling wave is an unwanted and generic feature of any numerical simulation we run the codes until a relaxation time is reached before the binary coalescence. We then reset the time and select another Bondi frame at rest with the center of mass at that time.
\subsection{The numerical setup}
We first perform the numerical integration of $\dot{M}$ (\ref{Mpunto}), $\dot{P}^i$ (\ref{Ppunto}), $\dot{D}^i$ (\ref{Dpunto}) and $\dot{J}^i$ (\ref{Jpunto}) where the initial values for those variables are taken from the metadata repository. The unwanted solitary waves that are present in any numerical integration are subtracted by finding the relaxation time (time where the wave is completely gone) and then subtracting the contribution of the traveling wave to the gravitational radiation at null infinity. Essentially this amounts to reset the initial time to the relaxation time and then use the Rochester data as the initial data for the evolution equations. Our Bondi system is such that at the relaxation time the origin of our coordinates is at the center of mass which is at rest at that time.The initial time is then reset to zero.

We then algebraically solve for 
\begin{align}
MR^i+c^{-2}\epsilon^{ijk}\frac{P^j}{M}S^k&=D^i+\frac{8}{5\sqrt{2}c} P^{j}\Delta\sigma _{R}^{ij}\nonumber\\
+&\frac{c^2}{G}\epsilon^{ijk}(\frac{4}{5}\sigma_I^{jl}\sigma_R^{kl}-\frac{36}{7}\sigma_I^{klm}\sigma_R^{jlm})\label{Rdot}
\end{align}
and
\begin{align}
	S^{i}+\epsilon^{ijk}R^j P^k=J^{i}+\frac{151c^3}{168\sqrt{2}G}(\sigma_I^{jk}\sigma_R^{ijk}-\sigma_I^{ijk}\sigma_R^{jk}).\label{S}
\end{align}

Where we have kept a first order expression of $V^i$ from the Bianchi identity of $\dot{D}^i$ (\ref{Dpunto}).
\section{Results}\label{section4}

Following the conventions of the Rochester repository, we have divided the numerical outcomes in different classes according to the relative masses into equal or non equal masses (EM or NEM). Likewise, the spins of the black holes are taken into account and we can distinguish between non spinning (NS), Aligned with the orbital angular momentum(A) or precessing spins (P). Thus an EM-P simulation has two black holes with equal mass and precessing spins non aligned with the orbital angular momentum.

\subsection{Data Analysis}
\subsubsection{A relationship between the initial angular momentum and the kickback velocity}
We plot below the relationship between  the initial angular momentum and the final velocity for two classes of initial data, EM and NEM. 

\begin{figure}[htp]
     \begin{subfigure}{0.5\textwidth}
         \centering
         \includegraphics[width=\textwidth]{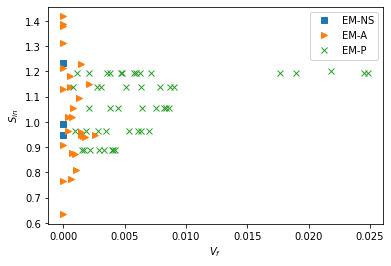}
         \caption{Final velocities for the EM class.}
         \label{fig3:f1}
     \end{subfigure}\hfill
     \begin{subfigure}{0.5\textwidth}
         \centering
         \includegraphics[width=\textwidth]{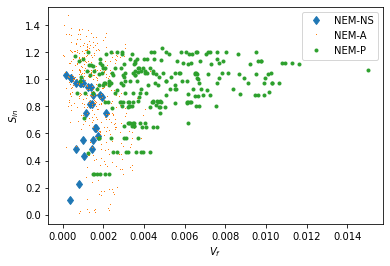}
         \caption{Final velocities for the NEM class.}
         \label{fig3:f2}
     \end{subfigure}
     \caption{Initial Angular momentum vs final velocity}
\end{figure}

We can see a clear difference between the precessing and non precessing cases in each class. The precessing binaries attain the highest final velocities although highly spread for the same values of initial angular momentum. The reason for the wide range of final velocities in the P cases lies in the intrinsic chaotic behavior of the precessing BHs.

One can see that there are only three simulations in figure \ref{fig3:f1} for the EM-NS case. The center of mass in those simulations does not acquire velocity as one might have guessed. The NEM-NS cases yield small and non chaotic final kicks. For the aligned cases the final velocities are small and densely distributed, showing a non chaotic behaviour.

It's worth mentioning figures \ref{fig3:f1} and \ref{fig3:f2} seem to achieve a peak for the final velocities in some range of initial $S_{in}$ depending on the classes. Nevertheless, this effect could be a bias due to the much less number of simulations for the values $S_{in}>1.2$.
\subsubsection{A relationship between the magnitudes of the initial and final angular momenta}

\begin{figure}[h]
	\centering
	\includegraphics[width=0.5\textwidth]{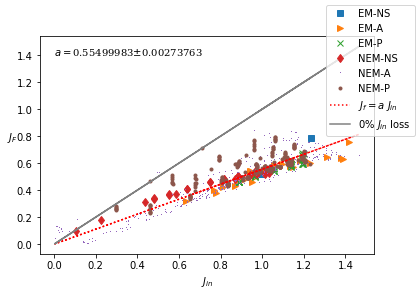}
	\caption{A relationship between the magnitudes of the initial and final angular momenta}
	\label{fig4}
\end{figure}

The loss of orbital angular momentum shows up for both the EM-NS, and the NEM-NS cases. The NEM-A class has a wide range of outcomes and in some cases the final angular momentum increases when the initial angular momentum is very close to zero. This special feature occurs when the spins are antialigned with the orbital angular momentum. Whereas the magnitude of the orbital part always decreases, the total angular momentum goes up since the spins remain approximately constants.
We made a fit of the final values obtained for all the classes to get a representative value of the outcomes for the simulations. This should not be regarded as the exact relation between $J_{in}$ and $J_f$ which is described by equation \ref{Jpunto}.
\subsubsection{The change in direction of the intrinsic angular momentum}\label{titlangle}
\begin{figure}[h]
	\centering
	\includegraphics[width=0.5\textwidth]{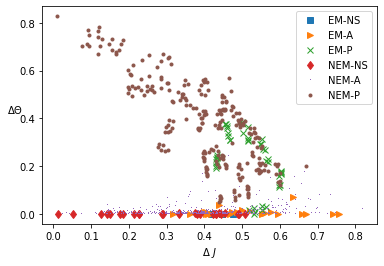}
	\caption{A relationship between the change in the magnitude of the angular momenta and the tilting angle}
	\label{fig5}
\end{figure}

To calculate the change in direction of the angular momentum we use the usual scalar product between the initial vector $J_{i}$ and the final $J_{f}$:
\begin{equation}
    \Delta \theta =\frac{\bold{J}_{f}\cdot \bold{J}_{i}}{|\bold{J}_{f}||\bold{J}_{i}|},
\end{equation}
where $\bold{J}_i$ in this particular case is the angular momentum at the relaxation time and $\bold{J}_f$ is the angular momentum at the final time of the evolution. $\bold{J}_i$ is here chosen in such a way to avoid the contribution of the traveling wave to the tilting angle. Thus we use the following definition at relaxation time
\begin{equation}
    \bold{J}_i=\bold{L}_{ni}+\bold{S}_{1i}+\bold{S}_{2i},
\end{equation}
where $\bold{L}_{ni}$,$\bold{S}_{1i}$,$\bold{S}_{2i}$ are the orbital angular momentum, the spin of the first and second components at relaxation time respectively. Hence, initial angular momentum is defined as Newtonian angular momentum since there is no radiation at that particular time. 
Given the above, fig. \ref{fig5} shows that NS systems have vanishing tilt. A similar situation occurs for the EM-A subclass whereas the NEM-A exhibit a small tilt. Overall, we see a tendency of the Aligned and Non-spinning configurations to have a tilting angle $\thicksim 0$, which we attribute to the known fact these configurations are stable ones \cite{kozameh2020spin}. 

On the contrary, preccesing systems are known to be chaotic. The most interesting cases are the P types which have a wide range of tilting angles. In particular, the tilting angle in NEM-P subclass appears to be inversely proportional to the change of angular momentum, the smaller the change in magnitude of S, the bigger the tilt. If the EM-P exhibit a similar behaviour is unknown due to the lack of simulation in the same range as NEM-P.

Finally we remark that, despite we calculate the change of the tilting angle using the total angular momentum at the relaxed and final time, it is important to note that the tilting angle is the same for the intrinsic angular momentum $S$. Indeed, at those times $\sigma \thicksim 0$ as there is no radiation at relaxation time and after the coalescence. Further, $\bold{R}\times \bold{P}\thicksim 0$ for the plot order in fig. \ref{fig11}, making $J_i=S_i$ and $J_f=S_f$.

\subsubsection{Correlation between the gravitational kick and the final value of $P^i/M$}
After selecting the new Bondi frame we run the codes using the values for M and S at the relaxation time and vanishing position and velocity for the center of mass.
Since the center of mass position is obtained algebraically and then numerically derived to obtain the kickback velocity it is worthwhile to check its relationship  with the final value of $P^i/M$. 

\begin{figure}[htb]
	\centering
	\includegraphics[scale=0.5]{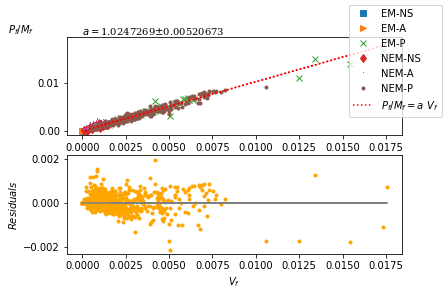}
	\caption{Correlation between $V_f$ and $(P^i/M)_f$.}
	\label{fig6}
\end{figure}

Note that the slope is $a=1.02$. This $2\%$ difference can be attributed to the fact that the numerical evolutions for $P^i$ and $M$ are affected by the linear drifts in $\sigma$, making a small contribution after the coalescence has taken place \ref{SigmaCalculus}.
\subsubsection{Relation between the radiation reaction force ($F_R$) and the gravitational radiation}
It is relevant to plot the time evolution of the (magnitude of the) radiation reaction force and its relation with the different modes of $\Psi_4$. To do that we define,

\begin{equation}
	F^i_R = \frac{1}{\sqrt{2}}\frac{d(M V^i)}{du}.
\end{equation}

and plot its time evolution of the different cases. In this case we select the simulation number 0443 but a similar pattern is obtained by any of the cases obtained. We first plot the contributions of the different modes of $\Psi_4$, for the whole process and zooming in the coalescence interval to enhance the differences between the modes

\begin{figure}[h]
	\begin{subfigure}{0.5\textwidth}
	    \centering
		\includegraphics[width=\textwidth]{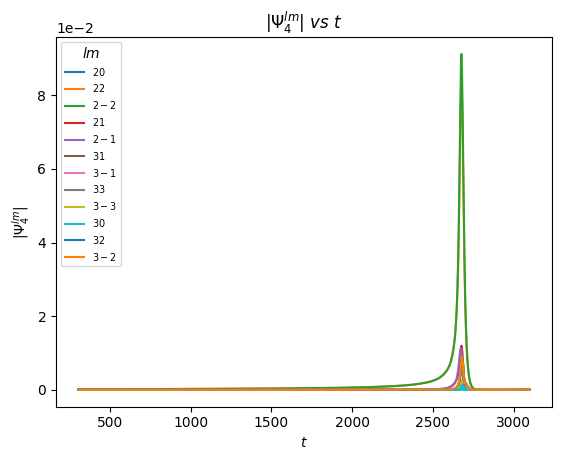}
		\caption{Whole evolution of the different modes}
		\label{fig7:f1}
	\end{subfigure}
	\hfill
	\begin{subfigure}{0.5\textwidth}
		\includegraphics[width=\textwidth]{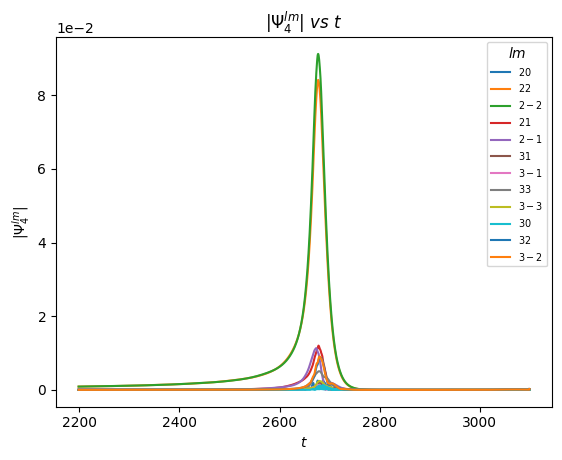}
		\caption{Zooming in to enhance the differences}
		\label{fig7:f2}
	\end{subfigure}
	\caption{Time evolution of the different modes of $\Psi_4$}
\end{figure}

As we can see in these graphs, each mode give a different contribution and they even peak at different times. Thus, depending on the coalescing scenario, the magnitude and direction of the impulse given by the $F_R$ can be quite different.

\begin{figure}[h]
	\begin{subfigure}[b]{0.5\textwidth}
		\includegraphics[width=\textwidth]{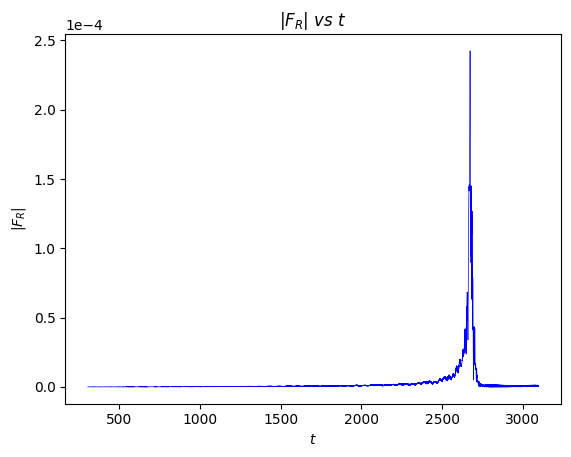}
		\caption{Whole evolution of $F_R$}
		\label{fig8:f1}
	\end{subfigure}
	\hfill
	\begin{subfigure}[b]{0.5\textwidth}
		\includegraphics[width=\textwidth]{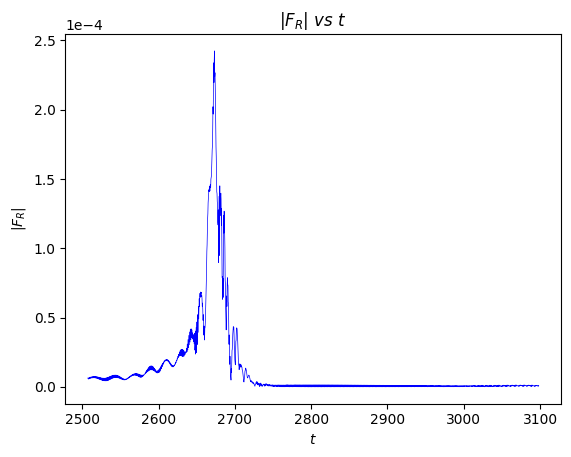}
		\caption{Zooming in to enhance the peak}
		\label{fig8:f2}
	\end{subfigure}
	\caption{Time evolution of the magnitude of $F_R$}
\end{figure}

In figure \ref{fig8:f2} we plotted the magnitude of $F_R$ for the $0443$ case. In this case the $F_R$ and the $l=2$ modes of the gravitational radiation peak at the same time but this can only be seen in the numerical data.

\subsubsection{Some typical behaviours of the center of mass motion while emitting gravitational radiation}
Finally, we plot the trajectory of the center of mass for selected representatives of each subclass. 

Note the difference in the scale of both plots. In the EM-NS case the center of mass remains virtually still and its final velocity is almost zero (1.3e-13 or 0.039 mm/s ). The NEM-NS is completely different, in this case the center of mass describes a spiral trajectory until the coalescing time where it is ejected at a final speed of 5.5e-04 or 165 km/s.
In the EM-A case the center of mass remains virtually in the same place and after emitting radiation remains at rest (1.26e-12 or 0.378 mm/s).In contrast to EM case, the center of mass in NEM-A case has a spiral trajectory that it is noticeable at the end and its final speed is 8.96e-04 (269 km/s).

Both configurations of equal masses (EM) when the spin is static, i.e. aligned or non-spinning, present a null kick velocity whereas the masses are non-equal the final kick velocities are noticeable. This suggests that equal masses configuration (A or NS) are likely to be stable and thus end up with no kickback and angular momentum in the same direction. 

In the precessing graphs, where the spins are not static, we observe a similar behaviour between the EM or NEM cases, even though the former is slightly lower than the latter. The final speeds for the EM and NEM cases are 6.66e-04 (200 km/s) and 1.18e-03 (354 km/s) respectively. We see the precessing spin plays an important role on increasing the kickback velocity in fig. \ref{fig9:f5}. This is a general feature of the class, as can be seen in \cref{fig3:f1,fig3:f2}.

\begin{figure}[tb!]
    \centering
	\begin{subfigure}{0.45\textwidth}
		\includegraphics[width=\textwidth]{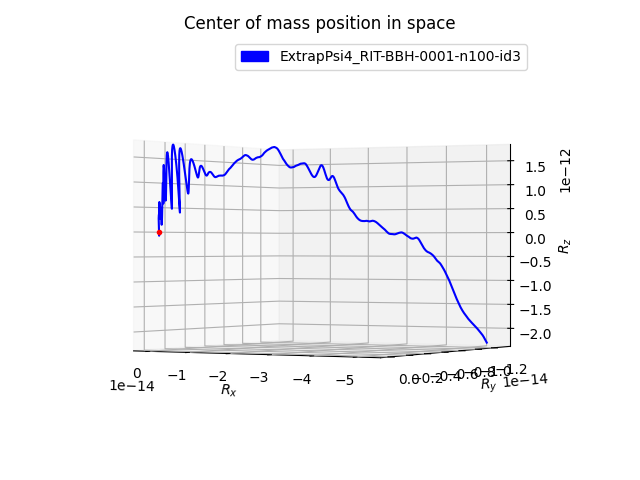}
		\caption{0001-EM-NS}
		\label{fig9:f1}
	\end{subfigure}\hfill
	\begin{subfigure}{0.45\textwidth}
		\includegraphics[width=\textwidth]{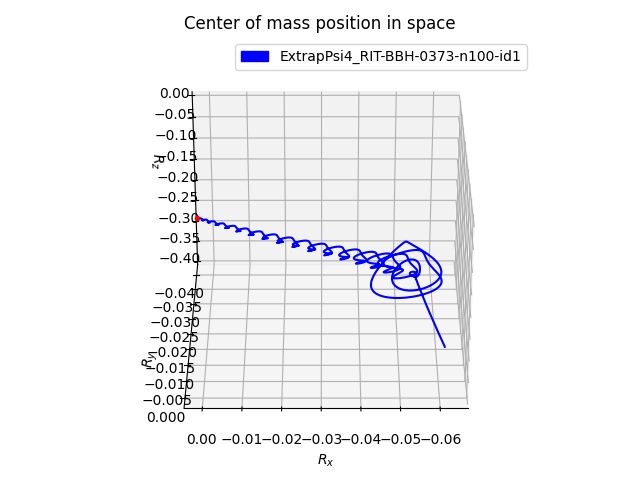}
		\caption{0373-NEM-NS}
		\label{fig9:f2}
	\end{subfigure}
\end{figure}
\begin{figure}[t]
    \centering
    \begin{subfigure}{0.45\textwidth}
		\includegraphics[width=\textwidth]{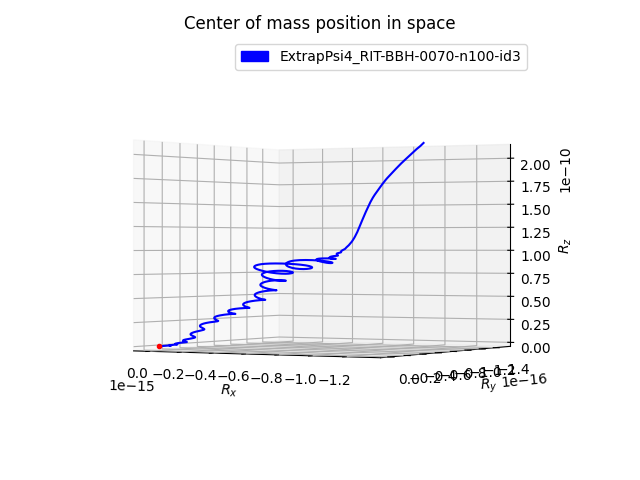}
		\caption{0070-EM-A}
		\label{fig9:f3}
    \end{subfigure}\hfill
	\begin{subfigure}{0.45\textwidth}
		\includegraphics[width=\textwidth]{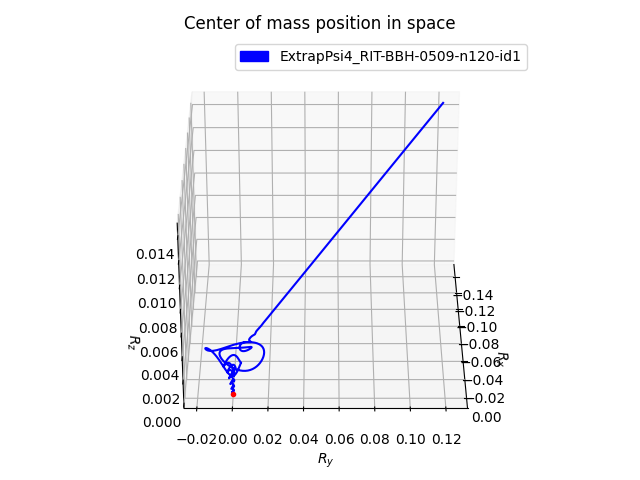}
		\caption{0509-NEM-A}
		\label{fig9:f4}
	\end{subfigure}
\end{figure}
\begin{figure}[tb!]
    \centering
	\begin{subfigure}{0.45\textwidth}
		\includegraphics[width=\textwidth]{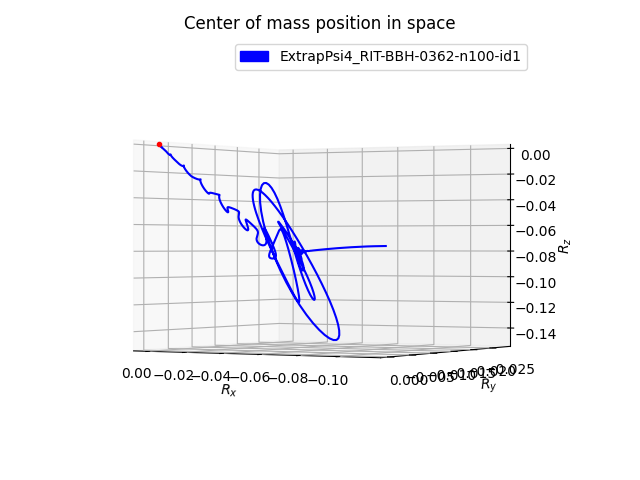}
		\caption{0362-EM-P}
		\label{fig9:f5}
	\end{subfigure}\hfill
	\begin{subfigure}{0.45\textwidth}
		\includegraphics[width=\textwidth]{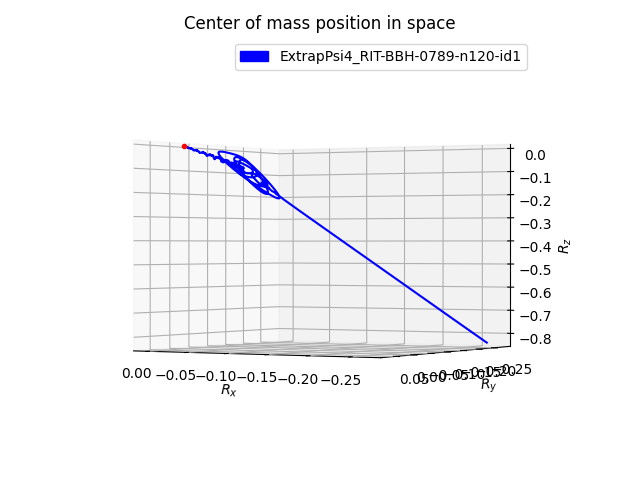}
		\caption{0789-NEM-P}
		\label{fig9:f6}
	\end{subfigure}
	\caption{Precessing spins.The red dot specifies the starting point of the center of mass and the number above the plot indicates the simulation from RIT catalogue where $\psi_4$ has been taken.}
\end{figure}

\subsection{Comparison with RIT catalogue Metadata}
There are some scalar quantities listed as final parameters in Rochester metadata simulations. In this section we compare Rochester parameters with the result obtained by \cref{Dpunto,Jpunto,Mpunto,Ppunto}. We show they are consistent except for some difference in decimals which we could be better matching making a more accurate evolution.

\subsubsection{Correlation between mass variations}
To compare the mass variation \ref{Mpunto} we defined the Rochester mass change as
\begin{equation}\label{deltaM_Roch}
    \Delta M_{Roch}=|M_F -M_{relax}|=M_{relax}-M_F,
\end{equation}
where $M_F$ is the final mass of the Kerr black hole and $M_{relax}$ is the mass at the relaxation time (which is slightly bigger than initial mass due to the traveling wave).

\begin{figure}[htb]
	\centering
	\includegraphics[scale=0.5]{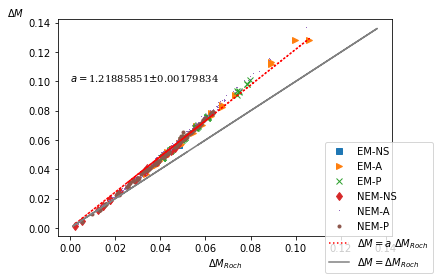}
	\caption{Correlation between $\Delta M$ and $\Delta M_{Roch}$.}
	\label{fig10}
\end{figure}

Plotting both $\Delta M$ leads to fig. \ref{fig10}. We see there is a 20\% difference from the ideal case. This could be attributed to the fact that defining \ref{deltaM_Roch} with $M_{relax}$, we are using a local measure of mass. This quantity should not differ too much from the asymptotic relaxed mass, as the only emitted radiation has been due to the travelling wave. Still, the effects discussed in \ref{SigmaCalculus} can be another cause of this discrepancy.
\subsubsection{Correlation between angular momentum changes}
\begin{figure}[htb]
	\centering
	\includegraphics[width=0.5\textwidth]{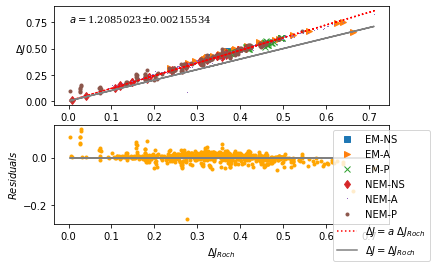}
	\caption{Correlation between $\Delta J$ and $\Delta J_{Roch}$.}
	\label{fig11}
\end{figure}

In figure \ref{fig11} we compare the variation $\Delta J$ obtained in the equation \ref{Jpunto} with the angular momentum variation in Rochester metadata which we define as
\begin{equation}\label{Jroch}
    \Delta J = |\chi M^2 -J_{ADM}|.
\end{equation}
The absolute value here makes $\Delta J$ positive. We are using as definition of final angular momentum that of a Kerr black hole $J_f=\chi M^2$, where $\chi$ and $M_f$ are the dimensionless spin parameter and the final mass of the resultant black hole respectively. As the components of the final angular momentum from the black hole are not provided, we use definition \ref{Jroch} which gives the correct change for non-spinning and aligned configurations. Yet this definition does not coincide exactly with the angular momentum change for precessing configurations.  

On the other hand, we subtract the absolute value of initial ADM angular momentum to estimate a change of the angular momentum. Note that making this subtraction we are considering the small contribution of the traveling wave to $\Delta J_{Roch}$ in fig.\ref{fig11}.

\begin{figure}[h]
	\centering
	\includegraphics[width=0.5\textwidth]{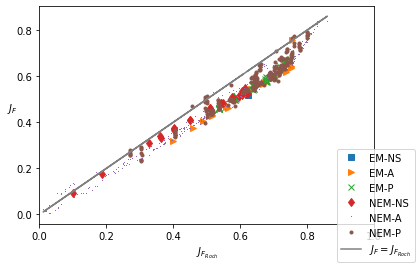}
	\caption{Correlation between the magnitude of final angular momenta}
	\label{fig12}
\end{figure}

 We end this section by showing the comparison of the obtained final angular momentum for the evolution equation \ref{Jpunto}. The correlation is shown in \ref{fig12}. $J_F$ is defined as
 \begin{equation}\label{Jnuestro}
     J_f=|\bold{J}_{F}|=|\bold{J}_{ADM}+\Delta \bold{J}| ,
 \end{equation}
 with $\Delta \bold{J}$ being that obtained with \ref{Jpunto}. Again, $J_{F_{Roch}}$ is defined as in \ref{Jroch}. As $\bold{J}_{ADM}$ is not the same for all the simulations, it is not trivial to expect a linear correlation of $J_F$ and $J_{Roch}$. Whereas a linear correlation is the desired result between both different formalisms, the reader should be aware that the methods for calculating the absolute value are different in each formalism. Indeed, eq. \ref{Jnuestro} is the absolute value of a vector while eq. \ref{Jroch} is the Kerr angular momentum formula obtained by matching the numerical evolution with a Kerr metric and its associated parameters.
\section{Final Comments and Conclusions}\label{section5}

We have numerically calculated the time evolution for the center of mass and intrinsic angular momentum for the 777 cases of BH coalescing binaries available in the Rochester repository. Since the final outcome of the process is a single black hole, the behaviour of $R^i$ and $S^j$ directly gives the position and spin of the final black hole.

To perform the numerical evolution we first derived a one to one correspondence between the spin weighted spherical harmonics of the gravitational radiation at the repository and the tensorial harmonics that are present in the equations of motion for $R^i$ and $S^j$.

We then addressed the issue of the unwanted traveling waves that are present in any initial data and removed them from the equations of motion for the 777 cases. This is done by first finding the relaxation time where the traveling wave is no longer present (amplitude smaller than the numerical error in the calculations). We then obtain the velocity of the center of mass at that time and perform a boost to a new Bondi frame whose origin coincides with the center of mass and it is initially at rest. For small traveling waves this translates into a time translation resetting the initial time to the relaxation time.

Using the equations of motion for the global variables available for asymptotically flat spacetimes we numerically obtain the time evolution for D, J, M and P and then use these variables to compute R and S.

The outcomes have been divided into two classes for the masses, equal mass (EM) or non equal mass (NEM), and three subclasses for the spins: non spinning (NS), spinning but aligned with the initial orbital angular momentum (A), and non aligned spins (P).

Some interesting cases are obtained and discussed. In particular, the subclass P of precessing cases exhibit a wide range of different outcomes for the final velocities whereas the NS and A subclasses do not. These appears to be related to the chaotic behaviour of the P subclass although a more thorough study is in order.
Likewise, the tilting angles of the P subclass have a non-ergodic distribution reaching up to $50 °$. This could also be attributed to the chaotic behaviour of spinning black holes \cite{kozameh2020spin,wu2015,levin2006}. The $F_R$ is defined and obtained for the 777 cases. Although none of the resulting BHs here analysed have relativistic speeds, one could easily give a relativistic definition by simply adding a gamma factor to the equation. A quantitative description of the trajectories is also given for representative members of each subclass.

Finally, the change in the Bondi Mass and angular momentum are compared between our approach and those from the Rochester catalog. By doing so we single out the gravitational radiation contribution and thus, introduce less potential sources of errors. The graphics show a clear correlation between the two approaches with a slope of $1.2$ in both graphs. We leave for future work to understand the source of this $20\%$ departure from the ideal correlation.

\begin{acknowledgments}
This research has been supported by grants from CONICET and the Agencia Nacional de Ciencia y Tecnolog\'ia.
\end{acknowledgments}

\nocite{*}

\bibliography{apssamp}
\end{document}